\documentclass[twocolumn,english,aps,prd,reprint,floatfix,notitlepage,footinbib,preprintnumbers,superscriptaddress,altaffilletter]{revtex4-2}

\usepackage{amsmath,amssymb,amsfonts}
\usepackage{hyperref,breakurl,cleveref,url}
\usepackage{color}

\usepackage{graphicx}
\usepackage[export]{adjustbox}

\usepackage{accents,mathrsfs,mathtools,dsfont}

\renewcommand{\paragraph}[1]{%
    \textit{#1}.---%
}

\def\skip{\vskip1.5pt}
\newcommand\trick[1]{}

\usepackage{enumitem}
\setlist[enumerate]{
    label={},
    leftmargin=2em,
    itemsep=2pt,
    topsep= 2pt,
    partopsep=0pt,
    parsep=0pt,
}

\usepackage{hyphenat}
\let\oldeqref\eqref
\renewcommand{\eqref}[1]{Eq.\,\smash{\oldeqref{#1}}}
\newcommand{\eqrefs}[2]{Eqs.\,\smash{\oldeqref{#1}} and \smash{\oldeqref{#2}}}

\newcommand{\rcite}[1]{Ref.\,\smash{\cite{#1}}}
\newcommand{\rrcite}[1]{Refs.\,\smash{\cite{#1}}}

\newcommand{\fref}[1]{Fig.\,\ref{#1}}

\DeclareMathOperator{\sinc}{sinc}

\def\mem{\hspace{0.1em}}
\def\hem{\hspace{0.05em}}
\def\nem{\hspace{-0.1em}}
\def\hnem{\hspace{-0.05em}}
\def\hhem{\hspace{0.025em}}
\def\hhnem{\hspace{-0.025em}}

\def\a{\alpha}
\def\b{\beta}

\def\m{\mu}
\def\n{\nu}
\def\r{\rho}
\def\s{\sigma}
\def\k{\kappa}

\def\t{\tau}

\def\bpsi{\bar{\psi}}

\def\tF{\tilde{F}}

\def\mplus{{\mem+\mem}}
\def\mminus{{\mem-\mem}}

\def\mdot{{\mem\cdot\mem}}

\def\swedge{{\mem{\wedge}\,}}

\def\da{{\dot{\a}}}
\def\db{{\dot{\b}}}

\def\lsq{{
    \kern-0.037em
    \adjustbox{scale=0.919,valign=c}{$
        {
            \adjustbox{raise=-0.0855em}{$\lfloor$}
            \llap{\reflectbox{\rotatebox[origin=c]{180}{$\lfloor$}}}
        }
    $}
    \kern-0.04em
}}
\def\rsq{{
    \kern-0.04em
    \adjustbox{scale=0.919,valign=c}{$
        {
            \rlap{\reflectbox{\rotatebox[origin=c]{180}{$\rfloor$}}} 
            \adjustbox{raise=-0.0855em}{$\rfloor$}
        }
    $}
    \kern-0.037em
}}

\def\M{{\mathcal{M}}}

\def\mathe{{\scalebox{1.05}[1.01]{$\mathrm{e}$}}}
\def\sprime{{\mathrlap{\smash{{}^\prime}}{\hspace{0.05em}}}}

\def\i{{\iota}}

\def\vex{\vec{x}}

\newcommand{\cop}[1]{ \mathrlap{c'}\phantom{c}^{\kern0.24em#1} }

\usepackage{tikz}
\usetikzlibrary{calc} 
\usetikzlibrary{shapes.geometric} 
\usetikzlibrary{positioning} 
\usetikzlibrary{fit} 
\usepackage[a]{esvect} 
\tikzset{empty/.style = {inner sep = 0pt, outer sep = 0, minimum size = 0}}
\tikzset{b/.style = {inner sep = 2pt, outer sep = 2pt, minimum size = 12pt}}
\tikzset{s/.style = {inner sep = 2.5pt, outer sep =2.5pt, minimum size = 1pt, font = \small}}
\tikzset{w/.style = {inner sep = 1pt, outer sep = 2pt, minimum size = 12pt, anchor = west}}

\definecolor{sky}{RGB}{144,187,231}
\definecolor{OxyRed}{RGB}{190,70,62}
\definecolor{NitroBlue}{RGB}{91,122,239}
\definecolor{HydrogenLight}{RGB}{245,250,252}

\tikzset{
	line/.style = {draw, line width = 1.1pt, line cap = round, rounded corners = 0.2pt},
	bine/.style = {draw, line width = 1.1pt, line cap = round, rounded corners = 0.0pt, dotted, color=OxyRed},
	dine/.style = {draw, line width = 1.4pt, line cap = round, rounded corners = 0.0pt, double},
	D/.style = {below = -1.1pt, font = \footnotesize\bfseries\sffamily},
	U/.style = {above = -1.2pt, font = \footnotesize\bfseries\sffamily},
	L/.style = {left,  font=\footnotesize},
	R/.style = {right, font=\footnotesize},
	X/.style = {circle, draw=black, fill=HydrogenLight, inner sep=0pt, outer sep=0pt, minimum size=4.5pt, line width=1.1pt},
	Y/.style = {circle, draw=black, fill=NitroBlue, inner sep=0pt, outer sep=0pt, minimum size=4.5pt, line width=1.1pt}
}

\usepackage[export]{adjustbox}

\newcommand{\bb}[1]{\bigg(\,{#1}\,\bigg)}
\newcommand{\BB}[1]{\Big(\,{#1}\,\Big)}
\newcommand{\bigbig}[1]{\big(\mem{#1}\mem\big)}

\newcommand{\bbsq}[1]{\bigg[\,{#1}\,\bigg]}

\newcommand{\lrp}[1]{\left(\mem{#1}\mem\right)}

\def\O{\mathcal{O}}

\def\rambda{\bar{\lambda}}
\def\trambda{\smash{\tilde{\lambda}}}
\def\tmu{\tilde{\mu}}

\newcommand\eq[1]{\begin{align}#1\end{align}}
\newcommand\eqsplit[1]{\begin{align}\begin{split}#1\end{split}\end{align}}

\newcommand\bealign[2]{\begin{aligned}[#1]#2\end{aligned}}

\newcommand\subeqref[2]{\begin{subequations}\label{#1}\begin{align}#2\end{align}\end{subequations}}

\newcommand{\pb}[2]{{\{\hem{#1},{#2}\hem\}}}

\def\MT{\mathcal{MT}}

\def\K{\mathcal{K}}

\def\Cinfty{C^\infty\hnem}

\def\Kerr{{\smash{\text{$\kern-0.075em\sqrt{\text{Kerr\hem}}$}}}}
\def\ga{{\text{G2A}}}

\begin{document}

\title{
	The Kerr-Newman Two-Twistor Particle 
}

\author{Joon-Hwi Kim}
\affiliation{Walter Burke Institute for Theoretical Physics, California Institute of Technology, Pasadena, CA 91125}

\begin{abstract}
	An all-orders worldline effective action
	for Kerr-Newman black hole
	is achieved in
	twistor particle theory.
	Exact hidden symmetries are identified
	in self-dual backgrounds.
\end{abstract}

\preprint{CALT-TH 2026-011}


\bibliographystyle{utphys-modified}

\renewcommand*{\bibfont}{\fontsize{8.5}{8.7}\selectfont}
\setlength{\bibsep}{1pt}

\maketitle

\paragraph{Introduction}%
An important topic in the current gravitational physics literature has been
the construction of worldline effective actions
\cite{Goldberger:2004jt,Porto:2005ac,Levi:2015msa,Porto:2016pyg,Levi:2018nxp,Kalin:2020mvi}
of spinning black holes.
Recently, \rcite{njmagic}
has derived an exact first-order action of the Kerr black hole
and provided its explicit all-orders formula,
by using
massive twistor theory
\cite{Penrose:1974di,Perjes:1974ra,tod1977some}
and the tangent bundle formalism for geodesic deviation \cite{gde}.

\rcite{njmagic}'s derivation
of the Kerr action
provides the test-particle counterpart
of the Newman-Janis trick
\cite{njmagic,probe-nj}.
The Newman-Janis trick is
a solution-generating technique
that rederived the Kerr metric
by performing a complex-geometrical transformation on
the Schwarzschild metric
\cite{Newman:1965tw-janis}.
This trick then
facilitated the very historical derivation of the Kerr-Newman solution \cite{Newman:1965my-kerrmetric}.
According to a modern view \cite{nja},
the Newman-Janis trick encodes
that the Kerr solution
is diffeomorphic to
a pair of self-dual and anti-self-dual Taub-NUT instantons
when taken as a holomorphic saddle.

The present paper
generalizes
\rcite{njmagic}'s construction
to derive the first-order action of the Kerr-Newman black hole
coupled to background gravitational and electromagnetic fields.
Again, the systematic formalism of \rcite{gde} proves instrumental in obtaining explicit formulae for the action.
We identify a Hamiltonian system
that can represent the exact, all-orders dynamics of the Kerr-Newman black hole
in its point-particle effective theory.

It is found that the conclusions of \rcite{njmagic} are preserved.
Nonlinear Newman-Janis shift manifests in self-dual backgrounds.
A chiral (``googly'') formulation of the Kerr-Newman action is viable
for generic, non-self-dual external fields.
In perturbation theory,
the spin exponentiation
of same-helicity Compton amplitudes
\cite{Guevara:2018wpp,Guevara:2019fsj,aho2020,Johansson:2019dnu,Aoude:2020onz,Lazopoulos:2021mna}
is manifested to all graviphoton multiplicities.

We also derive the classical equations of motion
in self-dual backgrounds.
We confirm the complexified, spinning analog of equivalence principle proposed by \rcite{sst-asym}.
This implies
dynamical hidden symmetries
in the presence of Killing-Yano tensors,
generalizing \rcite{probe-nj}'s previous result
about the Kerr particle.

\skip
\paragraph{Geodesic Deviation in Einstein-Maxwell Geometry}%
First, we present the new application of \rcite{gde}'s tangent bundle formalism
for geodesic deviation.

Let $(\M,g)$ be a real-analytic pseudo-Riemannian four-manifold,
and let $F {\:\in\:} \Omega^2(\M)$ be a two-form.
Let $\nabla$ be the Levi-Civita connection.
Let $D$ be the covariant exterior derivative due to $\nabla$.

Let $N \in T(T\M)$ be the horizontal lift
of the tautological vector field of the tangent bundle $T\M$
with respect to $\nabla$.
This is the geodesic spray, i.e., the generator of geodesic deviation in $T\M$.

\begin{widetext}
We explicitly evaluate 
$(\i_Nd)^{\ell-1}\mem \i_NF = (\i_ND)^{\ell-1}\mem \i_NF$
for all integers $\ell \geq 1$.
The result is
\begin{align}
\begin{split}
\label{PQseq}
   (\i_ND)^{\ell-1}\mem \i_N F
   \mem\,=\,\mem
\begin{aligned}[t]
&
           (P_\ell)_\m\mem
           dx^\m
           +
           (\ell\mminus1)\mem
           (P_{\ell-1})_\m\mem
           \, Dy^\m
\\
&
       +
       \sum_{p=2}^{\lfloor{(\ell+1)/2}\rfloor}\kern-0.4em
       \sum_{\a \in \Omega_p\hnem(\ell+1)}\kern-0.6em
           \mathscr{K}_\a
       \lrp{\hem
       \begin{aligned}[c]
       &
           \bigbig{\hnem
               P_{\a_1-1}
               Q_{\a_2}\, {\nem\cdots\mem} Q_{\a_p}
           \hnem}
           _{\nem\m}
           \, dx^\m
       \\
       &
           +
           (\a_p\mminus2)
           \bigbig{\hnem
               P_{\a_1-1}
               Q_{\a_2}\, {\nem\cdots\mem} Q_{\a_{p-1}} Q_{\a_p - 1}
           \hnem}
           _{\nem\m}
           \, Dy^\m
           \nem
       \end{aligned}
       }
   \,.
\end{aligned}
\end{split}
\end{align}
\vspace{-1.0\baselineskip}
\end{widetext}
\phantom{.}

\vspace{-2.0\baselineskip}\noindent
This is a new calculation,
not presented in our previous works
\cite{gde,njmagic}.
Yet, our notations are the same.
Each $\a = (\a_1,\a_2,\cdots,\a_p) \in \Omega_p(\ell)$
is an ordered partition
such that
$\a_1 + \a_2 + \cdots + \a_p = \ell$
and $\a_i \geq 2$.
The coefficient $\mathscr{K}_\a$
for $\a \in \Omega_p(\ell)$
is a product of binomial coefficients:
\begin{align}
\label{coeffK}
   \mathscr{K}_\a
   \,=\,
   \bbsq{
       \prod_{i=1}^p
       \binom{
           \bigbig{
               \sum_{j=i}^p \a_j
           } \mminus 2
           \hem
       }{
           \a_i \mminus 2
       }
       \nem\nem
   }
   \,.
\end{align}
For any $\a = (\a_1,{\cdots},\a_p)$
such that $\min\{\a_1,{\cdots},\a_p\} < 2$,
we set $\mathscr{K}_\a = 0$.
The $P$- and $Q$-tensors \cite{gde} are
\begin{subequations}
\begin{align}
   \label{P,Q}
     (P_\ell)_\m
     &\,=\,
         F_{\r_1\m;\r_2;\cdots;\r_\ell}(x)\,
         y^{\r_1} y^{\r_2} {\hem\cdots\mem} y^{\r_\ell}
     \,,\\
	(Q_\ell)^\m{}_\s
	&\,=\,
		R^\m{}_{\r_1\r_2\s;\r_3;\cdots;\r_\ell}\hnem(x)\,
		y^{\r_1} y^{\r_2} y^{\r_3} {\hem\cdots\mem} y^{\r_\ell}
	\,,
\end{align}
\end{subequations}
while
$P_\ell$ and $Q_\ell$
are set to zero
for $\ell = 0$ and $\ell = 0,1$, respectively.

\begin{figure}[h]
	\includegraphics[width=1.0\linewidth]{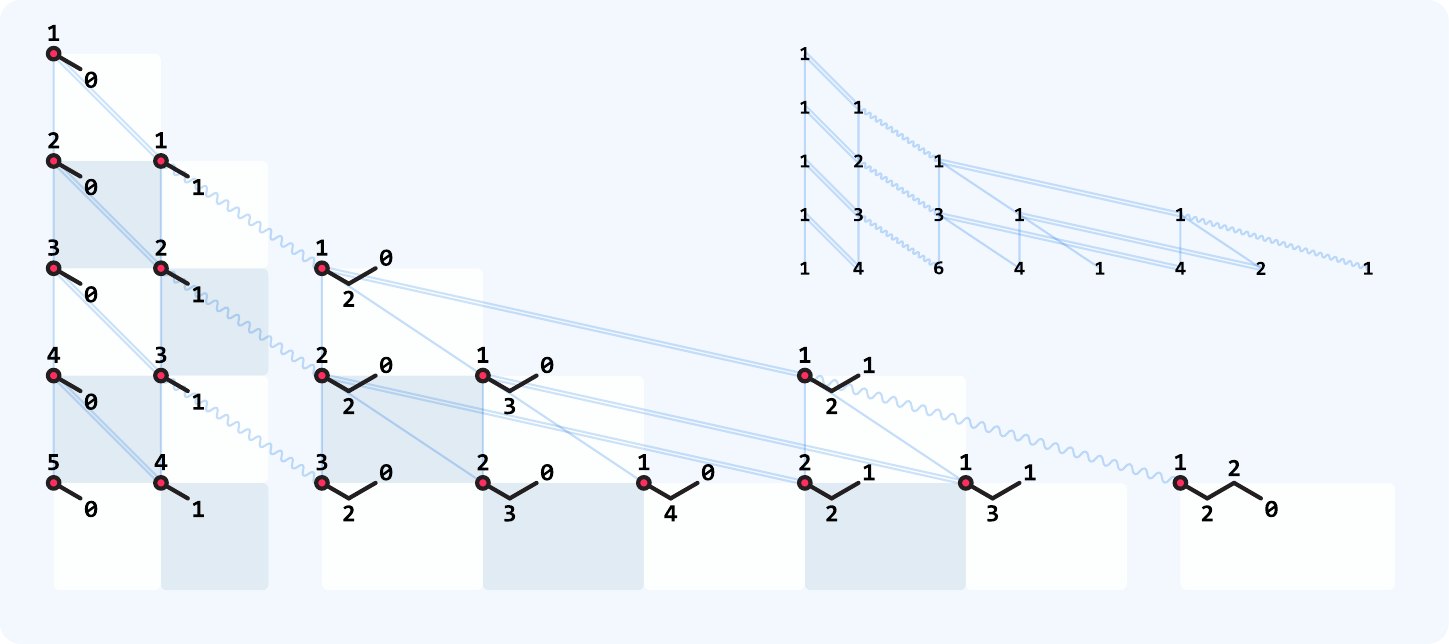}
	\caption{
		The ``organic chemistry'' of Einstein-Maxwell geometry.
		Molecules at each row follow from those on the former row by the action of $\i_ND$.
		Single lines represent the reaction
		$(P_{\a_1}Q_{\a_2}{\cdots\mem}Q_{\a_p})\mem dx \mapsto \sum_{\b_1+\cdots+\b_p = 1} (P_{\a_1+\b_1} Q_{\a_2+\b_2}{\cdots\mem}Q_{\a_p+\b_p})\mem dx$
		with $\b_1,{\cdots},\b_p {\:\geq\:} 0$.
		Double lines represent the reaction
		$(PQ{\mem\cdots\mem}Q)\mem dx \mapsto (PQ{\mem\cdots\mem}Q)\mem Dy$.
		Squiggly lines represent the ``polymerization'' reaction
		$(PQ{\mem\cdots\mem}Q)\mem Dy \mapsto (PQ{\mem\cdots\mem}Q)\mem (Q_2\hem dx)$.
		See \fref{molecules} and also \rcite{gde}.
		The smaller diagram in the upper right corner
		displays the coefficient of each molecule.
	}
	\label{chemistry-A}
\end{figure}

\eqref{PQseq}
is readily proven by mathematical induction,
based on the recursive structure illustrated in \fref{chemistry-A}.
This applies the ``organic chemistry'' intuition of \rcite{gde}.

Amusingly, \eqref{PQseq}
can be quickly seen by
a ``covariant color-kinematics'' \cite{cck} mapping,
which views gravity as a gauge theory of the Lorentz group.
Namely, \eqref{PQseq} can be deduced from
the formula for
$(\i_ND)^{\ell-1}\mem \i_N R^m{}_n$
given in \rcite{gde,njmagic}.
By viewing the antisymmetric pair of local Lorentz indices
as the adjoint index of the Lorentz group as a gauge group,
$(\i_ND)^{\ell-1}\mem \i_N R^m{}_n$
corresponds to $(\i_ND)^{\ell-1}\mem \i_N F^a$
where $F^a$ is a nonabelian field strength.
Taking the abelian limit then reproduces \eqref{PQseq}.

\skip
\paragraph{The Kerr-Newman Action}%
Second, we implement
the probe counterpart of Newman-Janis algorithm,
in the version of either
\rcite{njmagic} (twistorial, top-down)
or
\rcite{probe-nj} (relatively bottom-up),
to derive the exact effective action of the Kerr-Newman black hole
in the presence of external gravitational and electromagnetic fields.

The gravitational part has been already provided
by \rcite{njmagic}:
the Kerr two-twistor particle.
In this paper, we simply need to add the electromagnetic interaction action.
Let $A \in \Omega^1(\M)$ be the electromagnetic gauge potential,
and let $F = dA \in Z^2(\M) \subset \Omega^2(\M)$ now denote the electromagnetic field strength.

\begin{figure}[t]
	\centering
	\adjustbox{valign=c}{\begin{tikzpicture}
	    \node[empty] (O) at (0,0) {};
	    \node[empty] (X) at (3.75, 0) {};
	    \node[empty] (x) at (3.0, 0) {};
	    \node[empty] (Y) at (0, -0.85) {};
	    \node[w] (a00) at ($(O)$) {$A$};
	    \node[w] (a01) at ($(O)+1*(x)$) {$0$};
	    \node[w] (a10) at ($(O)+1*(Y)$) {$\i_NF$};
	    \node[w] (a11) at ($(O)+1*(Y)+1*(x)$) {$0$};
	    \node[w] (a20) at ($(O)+2*(Y)$) {$\i_ND\mem \i_NF$};
	    \node[w] (a21) at ($(O)+2*(Y)+1*(x)$) {$0$};
	    \node[w] (a30) at ($(O)+3*(Y)$) {$\vdots$};
	    \node[w] (a2K) at ($(O)+2*(Y)-1*(X)$) {$\i_N {\ast F}$};
	    \node[w] (a3K) at ($(O)+3*(Y)-1*(X)$) {$\i_ND\mem \i_N {\ast F}$};
	    \node[w] (a4K) at ($(O)+4*(Y)-1*(X)$) {$\vdots$};
	    \node[w] (a2k) at ($(O)+2*(Y)-1*(X)+(x)$) {$0$};
	    \node[w] (a3k) at ($(O)+3*(Y)-1*(X)+(x)$) {$0$};
	    \node[w] (phantom-a00) at ($(O)$) {$\phantom{\big|}$};
	    \node[w] (phantom-a01) at ($(O)+1*(x)$) {};
	    \node[w] (phantom-a10) at ($(O)+1*(Y)$) {};
	    \node[w] (phantom-a11) at ($(O)+1*(Y)+1*(x)$) {};
	    \node[w] (phantom-a20) at ($(O)+2*(Y)$) {};
	    \node[w] (phantom-a21) at ($(O)+2*(Y)+1*(x)$) {};
	    \node[w] (phantom-a30) at ($(O)+3*(Y)$) {};
	    \node[w] (phantom-a31) at ($(O)+3*(Y)+1*(x)$) {};
	    \node[w] (phantom-a2K) at ($(O)+2*(Y)-1*(X)$) {};
	    \node[w] (phantom-a3K) at ($(O)+3*(Y)-1*(X)$) {};
	    \node[w] (phantom-a4K) at ($(O)+4*(Y)-1*(X)$) {};
	    \node[w] (phantom-a2k) at ($(O)+2*(Y)-1*(X)+(x)$) {};
	    \node[w] (phantom-a3k) at ($(O)+3*(Y)-1*(X)+(x)$) {};
	    \draw[->] (a00)--(a01) node[midway,above] {\scriptsize \smash{${d}\mem\i_N$}\vphantom{d}};
	    \draw[->] (a10)--(a11) node[] {};
	    \draw[->] (a20)--(a21) node[] {};
	    \draw[->] (phantom-a00)--(phantom-a10) node[midway,left] {\scriptsize $\i_N{d}$};
	    \draw[->] (phantom-a10)--(phantom-a20) node[] {};
	    \draw[->] (phantom-a20)--(phantom-a30) node[] {};
	    \draw[<->] (phantom-a10)--(a2K) node[midway,above] {\adjustbox{raise=1.5pt}{\scriptsize dual}};
	    \draw[->] (phantom-a2K)--(phantom-a3K) node[] {};
	    \draw[->] (phantom-a3K)--(phantom-a4K) node[] {};
	    \draw[->] (a2K)--(a2k) node[] {};
	    \draw[->] (a3K)--(a3k) node[] {};
	\end{tikzpicture}}
	\caption{
		The ``$\pounds_N$ sequence'' for Einstein-Maxwell geometry.
	}
	\label{liecd-A}
\end{figure}

The ``$\pounds_N$ sequence'' for $A$ is shown in \fref{liecd-A}.
The recipe of \rrcite{njmagic,probe-nj} applies Hodge duality $\ell$ times on the $2^\ell$-pole moment term in this sequence.
This determines the electromagnetic interaction symplectic potential.
Consequently, the full symplectic potential
of the Kerr-Newman particle
is
\begin{align}
	\label{NJA}
	\theta_\text{KN}
	\,=\,
		&
	   	\BB{
	   		\cos(\pounds_N)
	   		+
	   		\sinc(\pounds_N)\,
	   		J\mem \i_N\hem d
	   	}\bigbig{
	   		p_\m\hhem dx^\m
	   	}
	   	\\
		&
		+
	   	\BB{
	   		\cos(\pounds_N)
	   		+
	   		\sinc(\pounds_N)\,
	   		\i_N\mem {*}\mem d
	   	}\bigbig{
	   		A_\m(x)\mem dx^\m
	   	}
	   	\,.
\nonumber
\end{align}
For simplicity, we have set the electric charge $q$ to a unit value;
to reinstate it, rescale $A \mapsto qA$.
We are now working in the 
the massive correspondence space $\mathcal{K}$
defined in \rcite{njmagic}.
Due to limited space, we 
leave the details to \rcite{njmagic}.
Recall that $\mathcal{K}$ is equipped with the geometric data
$N \in \Gamma(T\K)$ and $J : T^*\K \to T^*\K$,
where $J^2 = -\mathrm{id}$.

\eqref{PQseq} facilitates the explicit evaluation of \eqref{NJA}:
\vspace{-1.3\baselineskip}
\begin{widetext}
\vspace{-1.25\baselineskip}
\begin{align}
\begin{split}
\label{KN}
	\theta_\text{KN}
    	\mem=\,
    	{}&{}
    		\theta_{(0)}
   		   	 	+ A_\m(x)\mem dx^\m
	\\
    	 	{}&{}
    	 	+\mem\hhem \sum_{\ell=2}^\infty
    	 		\frac{1}{\ell!}
    			\sum_{p=1}^{\lfloor{\ell/2}\rfloor}\kern-0.2em
    			\sum_{\a \in \Omega_p(\ell)}\kern-0.2em
    	 			\mathscr{K}_\a\mem
    	 		\,p_\m
    	 		\lrp{
    	 		\bealign{c}{
    	 			&
				\bigbig{\hnem
					({*}^\ell Q_{\a_1}\hnem)\mem Q_{\a_2}\, {\nem\cdots\mem} Q_{\a_p}
				\hnem}
				{}^\m{}_\s\, dx^\s
				\\
				&
				+ (\a_p\mminus2)\mem
				\bigbig{\hnem
					({*}^\ell Q_{\a_1}\hnem)\mem Q_{\a_2}\, {\nem\cdots\mem} Q_{\a_{p-1}} Q_{\a_p - 1}
				\hnem}
				{}^\m{}_\s\, Dy^\s
			\hnem
    	 		}}
    	 \\
   	 	{}&{}
		+\mem\hhem
			\sum_{\ell=1}^\infty
				\frac{1}{\ell!}\,
		    	 		\bb{\hnem
						\bigbig{\hnem
							{*}^\ell P_\ell
						\hnem}
						{}_\m\, dx^\m
						+
						(\ell\mminus1)\mem
						\bigbig{\hnem
							{*}^\ell P_{\ell-1}
						\hnem}
						{}_\m\, Dy^\m
		    	 		}
    	\\
    	 	{}&{}
    	 	+\mem\hhem \sum_{\ell=2}^\infty
    	 		\frac{1}{\ell!}
		   \sum_{p=2}^{\lfloor{(\ell+1)/2}\rfloor}\kern-0.4em
		   \sum_{\a \in \Omega_p\hnem(\ell+1)}\kern-0.6em
		       \mathscr{K}_\a
		   \lrp{
		   \begin{aligned}[c]
		   &
		       \bigbig{\hnem
		           ({*}^\ell P_{\a_1-1})\mem
		           Q_{\a_2}\, {\nem\cdots\mem} Q_{\a_p}
		       \hnem}
		       _{\nem\m}
		       \, dx^\m
		   \\
		   &
		       +
		       (\a_p\mminus2)
		       \bigbig{\hnem
		           ({*}^\ell P_{\a_1-1})\mem
		           Q_{\a_2}\, {\nem\cdots\mem} Q_{\a_{p-1}} Q_{\a_p - 1}
		       \hnem}
		       _{\nem\m}
		       \, Dy^\m
		   \end{aligned}
		   \hnem}
 	 \,,
\end{split}
\end{align}
\vspace{-0.55\baselineskip}
\end{widetext}
\phantom{.}

\vspace{-2.15\baselineskip}\noindent
Here, ${*}^\ell P_j$ and ${*}^\ell Q_j$
means to act on the Hodge star $\ell$ times
on the electromagnetic field strength $F$ and the Riemann tensor $R$
inside $P_j$ and $Q_j$.

Again, $\theta_{(0)}$ is defined in \rcite{njmagic}:
the covariantization of the free-theory symplectic potential
of the two-twistor particle.
For the reader's sake, we reproduce it below:
\eq{
\label{theta0}
	\theta_{(0)}
	\,=\,
	p_\m\hhem dx^\m
	+ i\mem y^{\da\a}\mem
	\BB{
		\lambda_\a{}^I\hem D\rambda_{I\da}
		{\,-\,}
		D\lambda_\a{}^I\mem \rambda_{I\da}
	}
	\,.
}

By incorporating the first-class constraints $\phi_0 = \frac{1}{2}\mem (p^2 {\,+\,} m^2)$ and $\phi_1 = -p\mdot y$
via Lagrange multipliers,
\eqref{KN}
concretely defines an all-orders first-order worldline action 
in which each term 
is fully and explicitly evaluated
as a manifestly covariant worldline operator.

It should be clear that
\eqref{KN} nicely reduces to the Kerr and {\Kerr} symplectic potentials
in \rcite{njmagic}.
Also, it is straightforward that
the nonabelian Kerr-Newman particle
is obtained from \eqref{KN} by simple replacements
$A_\m(x)\mem dx^\m \mapsto i\mem \bpsi_i\mem D\psi^i$
and
$(P_\ell)_\s \mapsto q_a\hem (P_\ell)^a{}_\s$
(see the appendix of \rcite{njmagic} for the relevant setup).

\skip
\paragraph{Nonlinear Newman-Janis Shift}%
Third, we impose self-duality condition on both
electromagnetic and Riemannian curvatures
to reveal the dynamical, nonlinear Newman-Janis shift:
${*}F_{\m\n} = i\hem F_{\m\n}$,
${*}R^\m{}_{\n\r\s} = i\hem R^\m{}_{\n\r\s}$.

To this end, we simply replace ${*}^\ell$ with $i^\ell$ in \eqref{KN}.
Recalling the explanations given in \rcite{njmagic},
it is seen that this replacement results in
\begin{align}
	\label{NJA.exp}
	\theta_\text{KN}
	\,\sim\,
		\mathe^{i\pounds_N}\BB{
			p_\m\hhem dx^\m
			\hem+\mem\hhem A_\m(x)\mem dx^\m
			\hem+\mem\hhem \theta_\ga
		}
   	\,,
\end{align}
where $\sim$ signifies equivalence modulo total derivative.
Specifically,
we have discarded
$i\mem \sinc(\pounds_N)\mem d(A_\m(x)\mem y^\m)$.

\eqref{NJA.exp}
is nothing but 
a minimally coupled charged scalar particle---%
i.e.,
the Reissner-Nordstr\"om particle---%
geodesically deviated into complex spacetime,
up to the ``Gilbert-to-Amp\`ere'' replacement \cite{njmagic}
implemented by $\theta_\ga$
which reflects a kinematical detail:
\eq{
	\label{g2a}
	\theta_\ga
	\,=\,
		i\mem y^{\da\a}\mem
			\BB{
				\lambda_\a{}^I\hem D\trambda_{I\da}
				{\,-\,}
				D\lambda_\a{}^I\mem \trambda_{I\da}
			}
		- i\mem p_\m Dy^\m
	\,.
}

Clearly, this incarnates the spirit of the Newman-Janis algorithm
\cite{Newman:1965tw-janis}
and its very historical application
\cite{Newman:1965my-kerrmetric}
to the Reissner-Nordstr\"om solution
for the derivation of the Kerr-Newman solution.

The exponentiated Lie derivative $\mathe^{i\pounds_N}$ in \eqref{NJA.exp}
describes the map $\gamma$ defined in \rcite{njmagic}
that brings us to the ``primed chart''
$(z^{\m\sprime},y^{\m\sprime},\lambda_{\a'}{}^I)$
on the (complexified) massive correspondence space $\K$.
With this understanding, \eqref{NJA.exp} boils down to
\begin{align}
	\label{KN+}
	\theta_\text{KN}^+
	\,=\,
		p_{\m'}\hhem dz^{\m\sprime}
		- 2i\mem \trambda_{I\da'}\mem y^{\da\sprime\a\sprime}\mem
			D\lambda_{\a'}{}^I
		+ A_{\m'}(z)\mem dz^{\m\sprime}
   	\,,
\end{align}
where we have discarded a total derivative
just as in \rcite{njmagic}.
Here, $z^{\m\sprime} = \delta^{\m\sprime}{}_\m\mem (
	x^\m + iy^\m + \Gamma^\m{}_{\r\s}(x)\mem y^\r y^\s + \O(y^3)
)$
are the holomorphic coordinates
of curved spinspacetime.

Reviving the analysis in \rcite{probe-nj},
one sees that the worldline perturbation theory due to \eqref{KN+}
manifests the same-helicity (positive-helicity, in particular) spin exponentiation
of the graviphotonic Compton amplitudes
at all multiplicities
and any massless species.

\skip
\paragraph{Heavenly Equations of Motion}%
The classical equations of motion due to
the symplectic potential in \eqref{KN+}
and the first-class constraints $\phi_0$ and $\phi_1$
are easy to derive
by the technique of covariant symplectic perturbations
\cite{spt-math}.
By fixing their Lagrange multipliers respectively as
$\k^0(\t) = 1$ and $\k^1(\t) = 0$,
we find
\eqsplit{
	\label{heom}
	\dot{z}^{\m\sprime} 
		\mem=\mem p^{\m\sprime}
	&\,,\,\,\,
	\frac{Dy^{\m\sprime}}{d\t}
		\mem=\mem 
			F^{\m\sprime}{}_{\n'\hnem}(z)\, y^{\n\sprime}
	\,,\\
	\frac{D\lambda_{\a'}{}^I}{d\t}
		\mem=\mem 
			0
	&\,,\,\,\,
	\frac{D\trambda_{I\a\sprime}}{d\t}
		\mem=\mem \trambda_{I\b\sprime}\,
			\tF^{\db\sprime}{}_{\da'\hnem}(z)
	\,.
}
This shows that
the motion of
our Kerr-Newman particle
describes
a complex geodesic worldline $z^{\m\sprime}$
in spinspacetime
on which the left-handed spin frame
$\lambda_{\a'}{}^I$
is covariantly constant
(parallel-transported),
while the right-handed spin frame $\trambda_{I\da'}$
and the spin length pseudovector $y^{\m\sprime}$
are transported according to the Lorentz force law
due to the dynamically Newman-Janis shifted field strength.

Based on the fact that
the left-handed spinor bundle is flat for the moment,
$D\lambda_{\a'}{}^I\nem/d\t = 0$
explicitly verifies 
the spinning, complexified analog of equivalence principle,
which was proposed by \rcite{sst-asym}
as the property characterizing spinning black holes.
Namely,
there is no anti-self-dual component in the spin precession,
in self-dual backgrounds.

When re-covariantized at $x^\m$ in spacetime
(i.e.,
transcribed to the defining chart of $\K$),
\eqref{heom}
yields
an all-orders extension of the
Bargmann-Michel-Telegdi \cite{Bargmann:1959gz}
and 
Mathisson-Papapetrou-Dixon 
\cite{Mathisson:1937zz,Papapetrou:1951pa,Dixon:1970zza}
equations
with all the multipole coefficients, both electromagnetic and gravitational, set to unity.
This is indeed the desired behavior for the Kerr-Newman particle.
Crucial point here is that
the unity of the gravitational multipole coefficients
is achieved because we have utilized geodesic deviation
in generating the spin-curvature couplings in the action.

Similarly,
it is also easily seen that the holomorphic coordinates $z^{\m\sprime}$ of 
the Kerr black hole's
curved spinspacetime are Poisson-commutative
in this self-dual limit:
\eq{
	\label{zz=0}
	\pb{\hem z^{\m\sprime}\hem}{z^{\n'}} \,=\, 0
	\,.
}

\skip
\paragraph{Hidden Symmetry}%
Fourth, let us suppose
the presence of
a Killing vector $X \in \Gamma(T\M)$
as well as
a Killing-Yano tensor $Y \in \Omega^2(\M)$.
These shall satisfy 
\eq{
	\label{KYT}
	\pounds_X g = 0
	\,,\quad
	\pounds_X F = 0
	\,,\quad
	Y^\m{}_\r\hem F^\r{}_\n
	\,=\,
		F^\m{}_\r\hem Y^\r{}_\n
	\,,
}
and $Y_{\m\n;\r} {\:=\:} Y_{[\m\n;\r]}$.
The last condition
describes a commuting property,
which holds
for the Kerr-Newman background
by the type-D property \cite{Hughston:1972qf}
and also for
its self-dual sector:
the self-dual charged Taub-Newman-Unti-Tamburino (NUT) solution
in which $F^\m{}_\n \propto Y^\m{}_\n / |\vex|^3$.
See \rcite{nja} and also \rcite{Adamo:2023fbj}.

As is well-known \cite{Hughston:1972qf},
the Reissner-Nordstr\"om particle,
governed by the equations of motion
$\dot{x}^\m {\:=\:} p^\m$, $Dp^\m\nem/d\t = F^\m{}_\n(x)\mem p^\n$,
enjoys
the conserved quantities
\subeqref{charges0}{
	\label{charge0.Q}
	\mathds{Q} \,&=\,
		p_\m\hem X^\m(x)
		+ \a(x)
	\,,\\
	\label{charge0.C}
	\mathds{C} \,&=\,
		- p_\m\mem Y^\m{}_\r(x)\mem Y^\r{}_\n(x)\, p^\n
	\,.
}
In \eqref{charge0.Q},
$\a \in \Cinfty(\M)$ is a scalar field such that
$d\a = -\i_X F$ \cite{Hughston:1972qf},
whose local existence is guaranteed
because $-d\i_X F = \i_X dF = 0$.
Concretely, one can take $\a = \i_X A$
if one is willing to make an explicit gauge choice such that $\pounds_X A = 0$.
To derive \eqref{charge0.C},
one notes that
the covariant precession of the vector $Y^\m{}_\n(x)\mem p^\n$
is the same as $p^\m$
by virtue of the commuting property in \eqref{KYT}.

\rcite{probe-nj}
has established the bold proposal that
exact (all-orders in spin and coupling) conserved charges
of the spinning black hole probe
directly arise by
dynamically Newman-Janis shifting
the conserved quantities of the non-spinning black hole probe,
in the self-dual sector.
By appreciating this idea,
we identify
the following exact conserved charges
for our Kerr-Newman probe:
\subeqref{charges}{
	\label{charge.Q}
	\mathds{Q} \,&=\,
		p_{\m'} X^{\m\sprime}(z)
		+ \a(z)
	\,,\\
	\label{charge.R}
	\mathds{R} \,&=\,
		p_{\m'} Y^{\m\sprime}{}_{\n'\hnem}(z)\mem y^{\n\sprime}
	\,,\\
	\label{charge.C}
	\mathds{C} \,&=\,
		- p_{\m'} Y^{\m\sprime}{}_{\r'\hnem}(z)\mem Y^{\r\sprime}{}_{\n'\hnem}(z)\mem \mem p^{\n\sprime}
	\,.
}
In the explicit gauge choice for the electromagnetic gauge potential,
one finds $\a(z) {\:=\:} A_{\m'\hnem}(z)\mem X^{\m\sprime}(z)$.
It is easy to verify that $\mathds{Q}$, $\mathds{R}$, $\mathds{C}$ in \eqref{charges}
are all conserved by the equations of motion in \eqref{heom}.
Note that $Dp^{\m\sprime}\nem/d\t = F^{\m\sprime}{}_{\n'\hnem}(z)\mem p^{\n\sprime}$.
The covariant precession behaviors 
exhibited by
$p^{\m\sprime}$,
$y^{\m\sprime}$,
and
$Y^{\m\sprime}{}_{\n'\hnem}(z)\mem p^{\n\sprime}$
are all identical.

The conserved charges in \eqref{charges}
are sufficient to establish
\textit{integrability} of the Kerr-Newman probe dynamics
in self-dual Einstein-Maxwell backgrounds
admitting at least two Killing vectors and a Killing-Yano tensor.
In \rcite{probe-nj},
such integrability
has been identified as
the property that characterizes black holes
among all possible massive spinning objects.
While \rcite{probe-nj}
has substantiated this proposal
by the Kerr particle,
this paper reports the same conclusion
for the Kerr-Newman particle.

In fact, one can establish \textit{superintegrability}
by incorporating the anti-self-dual Pleba\`nski two-forms
\cite{Plebanski:1977zz},
just as in \rcite{probe-nj}:
\eq{
	\label{charge.K}
	\mathds{K}_a \,&=\,
		p_{\m'} (\Sigma_a)^{\m\sprime}{}_{\r'\hnem}(z)\mem Y^{\r\sprime}{}_{\n'\hnem}(z)\mem \mem p^{\n\sprime}
	\,.
}
It should be also highlighted that
the symmetry algebra 
formed by $\mathds{Q}$, $\mathds{C}$, and $\mathds{K}_a$
is \textit{identical}
to the non-spinning case
up to complexification,
by virtue of the very important property 
that $\pb{z^{\m\sprime}}{z^{\n\sprime}} = 0$
(\eqref{zz=0})
which is characteristic of black hole spinspacetime.

In short, we identify superintegrability
in the dynamics of the Kerr-Newman black hole binary system
at the zeroth self-force order,
yet in the self-dual sector.

\skip
\paragraph{Curved Massive Twistor Space}%
Fifth,
we identify the ``mu variables''
to return to the massive twistor space.

To this end, we trivialize the anti-self-dual spin connection coefficients,
employ tetrad formalism,
and expand the holomorphic coframe 
around a flat background as
$e^{\da\sprime\a} = dz^{\da\sprime\a} + h^{\da\sprime\a}$
which may utilize the second Pleba\'nski coordinates \cite{plebanski1975some}.
Then \eqref{KN+} becomes
\begin{align}
	\label{NJA'}
	\theta_\text{KN}^+
	\,=\,
		- \trambda_{I\da'}\hem dz^{\da\sprime\a}\hem \lambda_\a{}^I
		- 2i\mem \trambda_{I\da'}\mem y^{\da\sprime\a}\mem
			d\lambda_\a{}^I
		+ \theta'_\text{KN}
   	\,.
\end{align}
We have discarded a total derivative, while
\eq{
	\theta'_\text{KN}
	\,=\,
		\BB{
			- \trambda_{I\da'}\hem h^{\da\sprime\a}{}_{\m'\hnem}(z)\hem \lambda_\a{}^I
			+ A_{\m'\hnem}(z)
		}\,
			dz^{\m\sprime}
	\,.
}
As a result, the same identification for the mu variables 
as in \rcite{njmagic}
are made:
\eqsplit{
	\label{muvariables}
	\mu^{\da\sprime\mem I}
	\,=\,
		z^{\da\sprime\a}\hem\hhem \lambda_\a{}^I
	\,,\quad
	\tmu_I{}^\a
	\,&=\,
		\trambda_{I\da'}
		\bigbig{
			z^{\da\sprime\a} {\hem-\mem} 2iy^{\da\sprime\a}
		}
	\,.
}
Eventually, we find
\eqsplit{
	\label{twistor.h}
	\theta^+_\text{KN}
	\,&=\,
	- \trambda_{I\da'}\hem d\mu^{\da\sprime\mem I}
	+ \tmu_I{}^\a\hem d\lambda_\a{}^I
	+ \theta'_\text{KN}
	\,,\\
	\omega_\text{KN}
	\,&=\,
	d\mu^{\da\sprime\mem I} \swedge d\trambda_{I\da'}
	+ d\tmu_I{}^\a \swedge d\lambda_\a{}^I
	+ d\theta'_\text{KN}
	\,.
}
\eqref{twistor.h} identifies the notion of curved massive twistor space
in terms of
a deformation of the symplectic structure on massive twistor space:
\eq{
	\bigbig{
		\MT
		,\mem
		\omega^\circ
	}
	\,\,\,\xrightarrow{\,\,\,\,\,\,}\,\,\,
	\bigbig{
		\MT
		,\mem
		\omega^\circ \mplus\mem\hhem d\theta'_\text{KN}
	}
	\,.
}
This deformation is via a $(1,0)$-perturbation $\theta'_\text{KN}$
at the level of symplectic potential
such that the Poisson commutativity of holomorphic spinspacetime coordinates \cite{sst-asym} is preserved as \eqref{zz=0}.
Physically,
this encodes that
the gravitoelectromagnetic interaction action
is supported on the Newman-Janis shifted worldline $z^{\m\sprime}$
for the self-dual (incoming positive-helicity \cite{bialynicki1981note,ashtekar1986note}) sector.

As remarked in \rcite{njmagic},
the above recipes are motivated by the practical use of our framework
in perturbation theory,
so other pathways toward the definition of curved massive twistor space
might exist.

\skip
\paragraph{Googly Formulation}%
Lastly,
our final task is to derive the chiral (``googly'') formulation of the Kerr-Newman action.
This computes the generic, non-self-dual action
described in \eqref{KN}
in terms of the complex, primed variables
$(z^{\m\sprime},y^{\m\sprime},\lambda_{\a'}{}^I)$
by considering the map $\gamma$
of \rcite{njmagic}.

\vspace{-1.0\baselineskip}
\begin{widetext}
To this end,
we revive the gymnastics in \rcite{njmagic}.
Define
\eq{
\label{strings}
	&
	\theta^\pm_1
	\,=\,
		\frac{1}{2}\,
		\BB{
			p_\m Dy^\m
			\pm
				y^{\da\a}\mem
				\BB{
					\lambda_\a{}^I\hem D\trambda_{I\da}
					{\,-\,}
					D\lambda_\a{}^I\mem \trambda_{I\da}
				}
		\nem}
	+
		\i_N F^\pm
	\,,
}
where the $\pm$ superscripts describe
the projections by $\frac{1}{2}\mem (1 \mp i\mem {*})$.
Then it follows that
\subeqref{wsforms}{
\label{wsform.KN}
	\theta_\text{KN}
	\,\,&=\,\,
		p_\m\hhem dx^\m + A_\m(x)\mem dx^\m
	+
		\frac{
			\mathe^{i\pounds_N} \mminus 1
		}{\pounds_N}
		\,\theta^+_1
	+
		\frac{
			\mathe^{-i\pounds_N} \mminus 1
		}{\pounds_N}
		\,\theta^-_1
	\,,\\
\label{wsform.RN+ia}
	\mathe^{i\pounds_N}\BB{
		p_\m\hhem dx^\m + A_\m(x)\mem dx^\m\hnem
	}
	\,\,&\sim\,\,
		p_\m\hhem dx^\m + A_\m(x)\mem dx^\m
	+
		\frac{
			\mathe^{i\pounds_N} \mminus 1
		}{\pounds_N}
		\,\theta^+_1
	+
		\frac{
			\mathe^{i\pounds_N} \mminus 1
		}{\pounds_N}
		\,\theta^-_1
	\,.
}
The difference between \eqrefs{wsform.KN}{wsform.RN+ia}---%
Kerr-Newman and ``Reissner-Nordstr\"om${}+{} iy$''---%
derives
\eqsplit{
	\theta_\text{KN}
	\,\mem&\sim\,\,
	\mathe^{i\pounds_N}\mem
	\bb{
		p_\m\hhem dx^\m
		+ 
		\frac{
			\mathe^{-2i\pounds_N} \mminus 1
		}{\pounds_N}
		\,\theta^-_1
	}
	\,,\\
	\,\,&=\,\,
		\mathe^{i\pounds_N}\mem
		\bb{
			\theta_{(0)}
			- i\mem p_\m Dy^\m
			+\mem\hhem
				\sum_{\ell=2}^\infty
					\frac{(-2i)^\ell}{\ell!}\mem
						p_\m
						\bigbig{\hnem
							(\i_N\hem D)^{\ell-2} \i_N\mem R^-{}^\m{}_\n
						\hhnem}\hem y^\n
			+\mem\hhem
				\sum_{\ell=1}^\infty
					\frac{(-2i)^\ell}{\ell!}\mem
						(\i_N\hem d)^{\ell-1} \i_N\mem F^-
		}
	\,,
}
which leads to
\begin{align}
\begin{split}
\label{KN.googly}
	\theta_\text{KN}
    	\mem\,\sim\,\,
    	{}&{}
   			p_{\m'}\hhem dz^{\m\sprime}
			- 2i\mem \trambda_{I\da'}\mem y^{\da\sprime\a\sprime}\mem
				D\lambda_{\a'}{}^I
	   	 	+ A_{\m'\hnem}(z)\mem dz^{\m\sprime}
	\\
    	 	{}&{}
    	 	+\mem\hhem \sum_{\ell=2}^\infty
    	 		\frac{(-2i)^\ell}{\ell!}
    			\sum_{p=1}^{\lfloor{\ell/2}\rfloor}\kern-0.2em
    			\sum_{\a \in \Omega_p(\ell)}\kern-0.2em
    	 			\mathscr{K}_\a\mem
   	 		\,p_{\m\sprime}\hem
    	 		\lrp{
    	 		\bealign{c}{
    	 			&
				\bigbig{\hnem
					Q_{\a_1}^-\hem Q_{\a_2}\, {\nem\cdots\mem} Q_{\a_p}
				\hnem}
				{}^{\m\sprime}{}_{\s'}\mem dz^{\s\sprime}
				\\
				&
				+ (\a_p\mminus2)\mem
				\bigbig{\hnem
					Q_{\a_1}^-\hem Q_{\a_2}\, {\nem\cdots\mem} Q_{\a_{p-1}} Q_{\a_p - 1}
				\hnem}
				{}^{\m\sprime}{}_{\s'}\hhem Dy^{\s\sprime}\mem
			\hnem
    	 		}}
    	 \\
   	 	{}&{}
		+\mem\hhem
			\sum_{\ell=1}^\infty
				\frac{(-2i)^\ell}{\ell!}\,
		    	 		\bb{\hnem
						\bigbig{\hnem
							P^-_\ell
						\hnem}
						{}_{\s'}\mem dz^{\s\sprime}
						+
						(\ell\mminus1)\mem
						\bigbig{\hnem
							P^-_{\ell-1}
						\hnem}
						{}_{\s'}\hhem Dy^{\s\sprime}
		    	 		}
    	\\
    	 	{}&{}
    	 	+\mem\hhem \sum_{\ell=2}^\infty
    	 		\frac{(-2i)^\ell}{\ell!}
		   \sum_{p=2}^{\lfloor{(\ell+1)/2}\rfloor}\kern-0.4em
		   \sum_{\a \in \Omega_p\hnem(\ell+1)}\kern-0.6em
		       \mathscr{K}_\a
		   \lrp{
		   \begin{aligned}[c]
		   &
		       \bigbig{\hnem
		           P^-_{\a_1-1}\mem
		           Q_{\a_2}\, {\nem\cdots\mem} Q_{\a_p}
		       \hnem}
		       _{\nem\m'}
		       \mem dz^{\m\sprime}
		   \\
		   &
		       +
		       (\a_p\mminus2)
		       \bigbig{\hnem
		           P^-_{\a_1-1}\mem
		           Q_{\a_2}\, {\nem\cdots\mem} Q_{\a_{p-1}} Q_{\a_p - 1}
		       \hnem}
		       _{\nem\m'}
		       \hhem Dy^{\m\sprime}
		   \end{aligned}
		   }
 	 \,.
\end{split}
\end{align}
\newpage
\end{widetext}
%

\noindent Again, we find the structure that
the first entry in each string of $P$ or $Q$ tensors
develops the anti-self-dual projection.
\eqref{KN.googly} reproduces \eqref{KN+}
in the self-dual limit.
As a friendly reminder, our index notation \cite{gde} implies
\begin{subequations}
\begin{align}
   \label{P,Q(z)}
     (P_\ell)_{\m'}
     &\,=\,
         F_{\r\sprime_1\m\sprime\hem;\r\sprime_2;\cdots;\r\sprime_\ell}(z)\,
         y^{\r\sprime_1} y^{\r\sprime_2} {\hem\cdots\mem} y^{\r\sprime_\ell}
     \,,\\
	(Q_\ell)^{\m\sprime}{}_{\s'}
	&\,=\,
		R^\m{}_{\r\sprime_1\r\sprime_2\s\sprime\mem;\r\sprime_3;\cdots;\r\sprime_\ell}\hnem(z)\,
		y^{\r\sprime_1} y^{\r\sprime_2} y^{\r\sprime_3} {\hem\cdots\mem} y^{\r\sprime_\ell}
	\,.
\end{align}
\end{subequations}

\eqref{KN.googly} provides
the googly formulation of the Kerr-Newman action.
To reiterate,
it describes the completely general case of
non-self-dual
gravitoelectromagnetic field configurations
but formulates the self-dual and anti-self-dual modes
on an unequal footing.
It provides a worldline action localized on the holomorphic worldline $z^{\m\sprime}$,
the perturbation theory of which
manifests the same-helicity spin exponentiation
for incoming positive-helicity photons and gravitons.
The photon coupling is always linear.
Graviphoton contact vertices arise iff the incoming photon carries negative helicity.
When examined in the spin length expansion,
the first graviphoton contact vertex
appears at $\O(y^3)$.

\skip
\paragraph{Interpretation}%
We end with quick remarks on the physical interpretation
by recalling the comments given in \rcite{njmagic}.
\eqref{KN} represents two charged masses
joined by 
a Dirac-Misner string
(line defect of magnetic and NUT fluxes).
\eqref{KN} represents 
a dangling anti-self-dual Dirac-Misner string
attached to
a charged mass.
The shape of the Dirac-Misner string
happens to be geodetic
for each slice of simultaneity defined by the worldline proper time.
For the Dirac string part,
this is completely justified
based on its well-established status
as a topological surface operator.

By virtue of the transparent decoding of the Newman-Janis algorithm
revealed by \rcite{nja},
we readily realize that
this string structure
is the precise infrared counterpart
of the actual Dirac-Misner string
that exists in the ultraviolet description of the Kerr-Newman solution
\footnote{
	Here,
	``infrared'' refers to the point-particle effective theory,
	while
	``ultraviolet'' refers to general relativity.
}.
Namely, it is shown in \rcite{nja}
that the Kerr-Newman solution
as a holomorphic saddle in Einstein-Maxwell theory
represents a pair of self-dual and anti-self-dual
charged Taub-NUT instantons.
In conclusion,
the holomorphic worldline $z^{\m\sprime}$
is the infrared avatar of the worldline of the self-dual charged Taub-NUT instanton
that constitutes the Kerr-Newman black hole.

The googly action in \eqref{KN.googly}
perturbs around the self-dual sector---%
the sector 
in which we, at least, have a full understanding on dynamics
in terms of remarkable properties
such as the superintegrability due to \eqrefs{charges}{charge.K}
or the same-helicity spin exponentiation.
The googly agenda means to approach the complicated dynamics of the Kerr-Newman probe
from its exactly solvable part,
which should sound reasonable.

Note that \rcite{gmoov} had speculated on a worldsheet structure,
although no clear physical interpretation
nor explicit formulae such as \eqrefs{KN}{KN.googly}
were provided.

\medskip
\paragraph{Acknowledgements}%
J.-H.K. is supported by the Department of Energy (Grant No.~DE-SC0011632) and by the Walter Burke Institute for Theoretical Physics.

\onecolumngrid

\medskip
\medskip{
\centering
\rule{0.70\textwidth}{0.2pt}\\[-1.0\baselineskip]
\rule{0.43\textwidth}{0.4pt}\\[-1.0\baselineskip]
\rule{0.32\textwidth}{0.7pt}\\[-1.0\baselineskip]
\rule{0.21\textwidth}{1.0pt}\\[-1.0\baselineskip]
\rule{0.10\textwidth}{1.3pt}\\[-1.0\baselineskip]
\vspace{1.0\baselineskip}
}\medskip

\newpage

\begin{figure}[h]
	\includegraphics[scale=0.85]{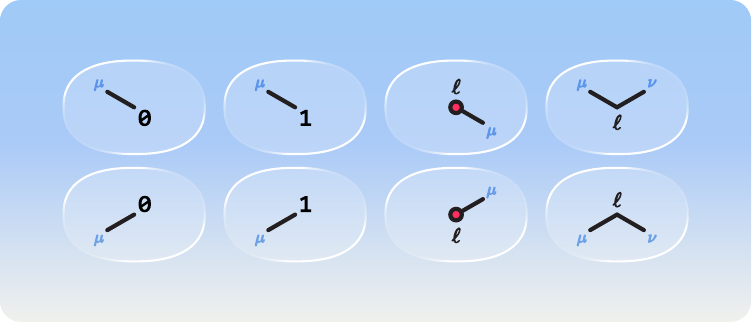}
	\caption{
		The graphical notation used in \fref{chemistry-A}.
		Elements in the first row denote
		$dx^\m$, $Dy^\m$, $(P_\ell)_\m$, and $(Q_\ell)^{\m\n}$,
		respectively.
		Elements in the second row denote
		$dx_\m$, $Dy_\m$, $(P_\ell)^\m$, and $(Q_\ell)_{\m\n}$,
		respectively.
	}
	\label{molecules}
\end{figure}

\twocolumngrid

\bibliography{references.bib}

\end{document}